\documentclass[aps,prl,twocolumn,groupedaddress,showpacs]{revtex4}
\usepackage{graphicx} 
\usepackage{bm} 
\usepackage{amssymb,amsmath}
\begin{document}

\title{The magnonic limit of  domain wall propagation in ferromagnetic
nanotubes}    \author{Ming     Yan}    \affiliation{Institut    f\"ur
Festk\"orperforschung   (IFF-9),   Forschungszentrum  J\"ulich   GmbH,
D-52428     J\"ulich,      Germany}     \author{Christian     Andreas}
\affiliation{Institut     f\"ur     Festk\"orperforschung     (IFF-9),
Forschungszentrum   J\"ulich    GmbH,   D-52428   J\"ulich,   Germany}
\author{Attila        K\'akay}       \affiliation{Institut       f\"ur
Festk\"orperforschung   (IFF-9),   Forschungszentrum  J\"ulich   GmbH,
D-52428    J\"ulich,   Germany}   \author{Felipe    Garcia-Sanchez   }
\affiliation{Institut     f\"ur     Festk\"orperforschung     (IFF-9),
Forschungszentrum   J\"ulich    GmbH,   D-52428   J\"ulich,   Germany}
\author{Riccardo       Hertel}       \affiliation{Institut       f\"ur
Festk\"orperforschung   (IFF-9),   Forschungszentrum  J\"ulich   GmbH,
D-52428     J\"ulich,     Germany}    \email[Corresponding     author:
]{r.hertel@fz-juelich.de} \date{\today}

\begin{abstract}
 We  report a study on the field-driven  propagation  of vortex-like  domain walls  in
ferromagnetic   nanotubes.    This
particular geometry gives rise to a special feature of the static wall
configuration,  which significantly  influences its  dynamics.  Unlike
domain walls in  flat strips, the left-right symmetry  of domain
wall  propagation is  broken. Furthermore,  the domain  wall velocity  is not
limited by the  Walker  breakdown.  Under  sufficiently large  magnetic
fields, the domain  wall velocity reaches the velocity of spin  waves (about 1000
$m/s$) and is thereafter connected with a direct emission of spin waves.  The
moving domain  wall maintains its main structure  but has characteristic
spin-wave  tails attached.   The spatial  profile of  this topological
soliton is determined by the spin-wave dispersion.

\end{abstract}

 \maketitle

The  dynamic properties  of objects  moving in  a medium  often change
dramatically  as soon  as their  velocity exceeds  the speed  of waves
propagating in  the medium. Well-known examples thereof  are the sonic
barrier encountered when an object propagates through air at the speed
of  sound,  or  the  Cherenkov  radiation \cite{cherenkov} which  occurs  when  charged
particles traverse a dielectric  medium at velocities above the phase
velocity of light. In both cases, the motion of the object is strongly
damped  above this  limit --  much  stronger than  below this  critical
velocity. If the object is accelerated, this critical velocity in fact
acts as a 'barrier' above which a large part of the object's energy is
radiated into the medium in form  of waves. For this reason, in the early times of aviation, the  speed of sound was representing a limit
of the  achievable velocity  of a plane.

In this letter we report on simulations predicting an analogous effect
in ferromagnetic nanostructures. The role of the medium is here played
by the  ferromagnetic material, the propagating object  is a
domain wall (DW), and the radiation above the critical velocity occurs
in the form of spin waves emitted by the DW. The magnonic limit of DW propagation was known in weak ferromagnets with a thickness in the $\mu m$ scale \cite{soliton_book} and analytically predicted for moving Bloch walls \cite{bouzidi}. We here propose a particular geometry of magnetic nanostructures, a cylindrical tube, in which indeed a magnonic barrier occurs. A rather unique manner of spin-wave emission by supermagnonic DWs is discovered.

One  major  difficulty  encountered  when  studying  such  effects  in
magnetic nanostructures  is the very  fast DW propagation  required to
reach the velocity of spin waves,  which is in the order of about 1000
$m/s$  for Permalloy (Py)  -- much  higher  than the  typical DW speed  in
nanostructures. Such high velocities are problematic because the speed
of  DW is  usually limited  by the  Walker breakdown \cite{walker},  a micromagnetic
instability occurring  at a much  lower critical velocity.  The Walker
breakdown is connected with a  structural change of the DW and usually
leads to an irregular, oscillatory DW motion\cite{walker, nakatani}.
Although it has recently
been reported that in cylindrical nanowires a special type of massless
DWs can develop which are not affected by the Walker limit \cite{yan}, we quickly
found that just because of the  absence of mass this type of DW does
not yield the desired interaction  with spin waves. We then found that
one could  obtain both, supermagnonic DW  velocities and non-vanishing
DW mass, in the case  of magnetic nanotubes.
In such a tubular geometry, we also discovered an unexpected dynamic behavior of the vortex-like DWs:  
the propagation of  these DWs
breaks the  left-right symmetry, which  to our knowledge is  taken for
granted in  any other case.  This particular behavior, which  will be
discussed in more  detail further below, leads to  the occurrence of a
favorable and unfavorable propagation direction --
or DW chirality, the latter  being just a different perspective on the
same effect.

\begin{figure}
\begin{center}
\includegraphics[width=3.4in]{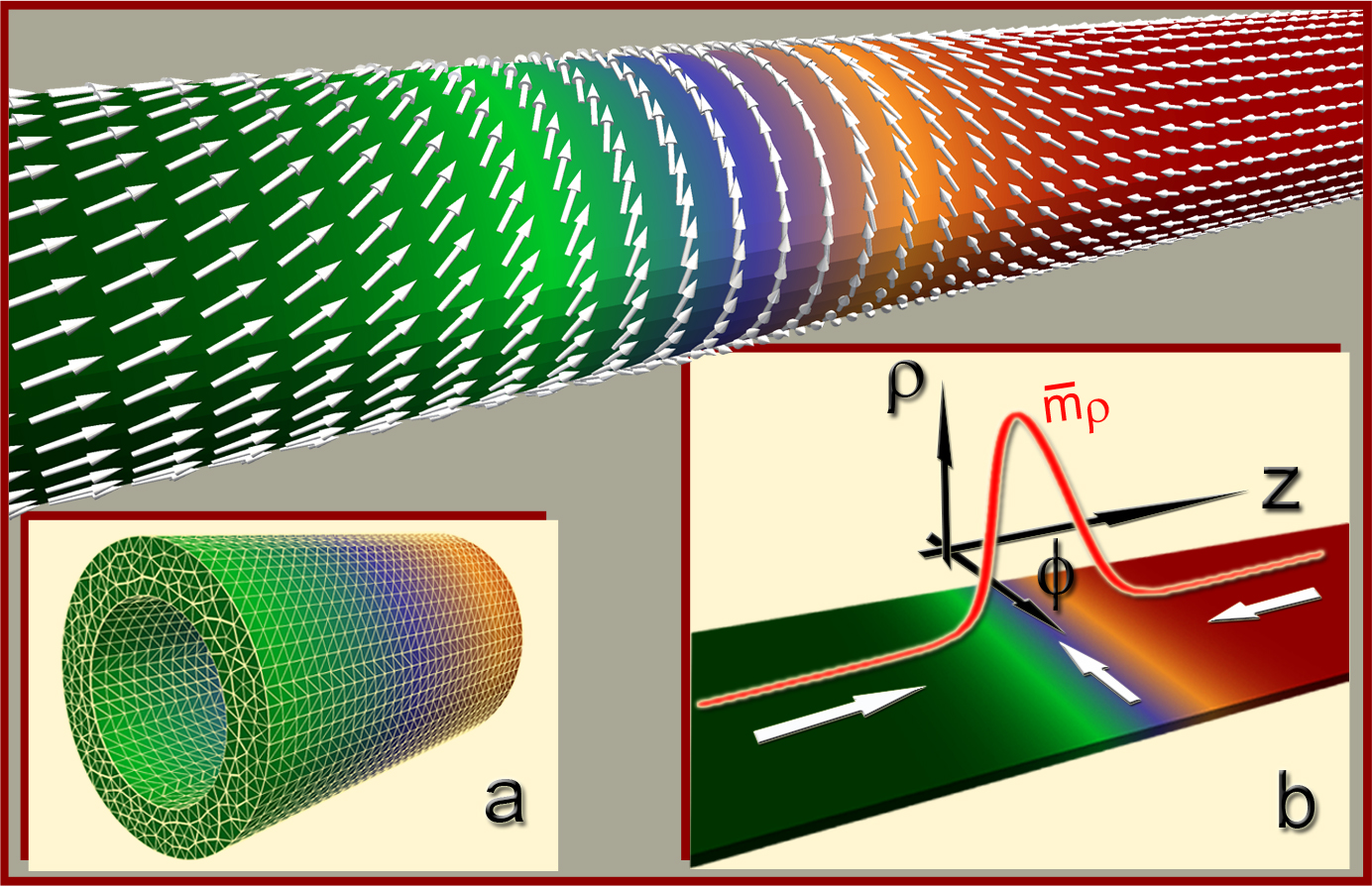}
\end{center}
\caption{\label{fig1}A vortex-like DW formed in a 4 $\mu m$ long Py tube
with 40~$nm$ inner diameter and 10~$nm$ thickness. (a) A small section
of the  tube. (b)  A flat strip  obtained from 'unrolling'  the curved
surface of  the tube with  a cylindrical coordinate system. The  white arrows
indicate the magnetization  near the DW. The red curve  is the plot of
$\bar{m}_{\rho}$  near   the  DW  indicating   a  non-zero  radial
component of the magnetization.}
\end{figure}

Similar to the case of thin  strips \cite{mcmichael}, different types of DWs
can  form in  cylindrical wires  \cite{hertel_jmmm02}. With  sufficiently large
diameter, a vortex-like DW  is energetically favorable, which contains
a Bloch point in the middle. The Bloch point can be avoided by using a
hollow cylindrical  wire (a tube)\cite{hertel_jmmm04, landeros}.  It is  worthwhile to notice
that  a nanotube  can be  considered as  a curved  thin strip  without 
lateral boundary.   As we will show  later, both  features of the
tube,  the  curvature of the surface  and  the periodic  boundary condition,  have  significant
influences   on  the   DW  dynamics.    Figure~\ref{fig1}   shows  the
configuration of a  vortex-like head-to-head (h2h) DW formed in a 4 $\mu m$ long 
Py tube with  40 $nm$  inner  diameter and  10 $nm$  thickness. The direction of the local magnetization is indicated by a unit vector $\vec{m}$ in a cylindrical coordinate system with the  $z$ axis along  the tube. In
the DW region,  the magnetization circles around the  tube and forms a
coreless vortex.
 Such a DW  is analogous to a transverse DW in  thin strips, which can
be displayed  by artificially 'unrolling'  the tube into a  flat thin
strip  as  shown  in  Fig.~\ref{fig1}b.  Obviously,  there  exist  two
energetically  degenerate  configurations  of  the  DW  with  opposite
vorticity ($\pm m_\phi$). This corresponds to the two equivalent orientations
of the transverse DWs along  the width direction in flat strips.  The
curvature of the tube, however, has an important  consequence on the DW
configuration.   Unlike a  transverse  DW in  flat  strips, where  the
magnetization lies perfectly in plane,  the vortex-like h2h DW in the
tubes  has  a small  positive  radial  component  of the  magnetization
($m_\rho$).  This is  illustrated by the plot of the averaged $m_\rho$  over
each cross  section of the  tube ($\bar{m}_{\rho}$) near the DW in Fig.~\ref{fig1}b.  The  presence of
the  non-zero $m_\rho$  is  to reduce  one  major source  of the  DW
energy,  the volume  charge generated  in  the DW  region. The  volume
charge density is proportional  to the divergence of the magnetization
field.    In cylindrical  coordinates,   the  divergence   of  the
magnetization     vector      $\vec{m}$     is     given      by     
\begin{equation}
\vec{\nabla}\cdot\vec{m}=\frac{m_\rho}{\rho}+\frac{\partial m_\rho}{\partial \rho}+ 
\frac{1}{\rho}\frac{\partial m_\phi}{\partial \phi}+
\frac{\partial m_z}{\partial z}.
\end{equation}
The
last  term is negative  for h2h  DWs and positive for tail-to-tail (t2t) DWs.  Clearly,  by having  a positive
$m_\rho$, the total volume charge of the h2h DWs is compensated.
 For  the   same  reason,  a  negative  $m_\rho$   should  appear  for
t2t  DWs  in nanotubes,  which  is confirmed in  our
simulations.

The   magnetization  dynamics   of  the   DW  is   described   by  the
Landau-Lifshitz-Gilbert equation :
\begin{equation}\label{eqn:llg}
\frac{d\vec{M}}{dt}   =  -\gamma\vec{M}\times \vec{H}_{\rm   eff}  +
\frac{\alpha}{M_s}\left[\vec{M}   \times  \frac{d\vec{M}}{dt}  \right],
\end{equation}
where  $\vec{M}$ is  the local  magnetization, ${M_s}$  the saturation
magnetization,  $\gamma$ the  gyromagnetic ratio,  $\vec{H}_{\rm eff}$
the  effective field,  $\alpha$ the  Gilbert damping  factor.   In our
simulations,   Eq.~\ref{eqn:llg}  is   solved   numerically  using   a
finite-element method \cite{hertel_jap01}.
Typical material  parameters of Py,  $\mu_0M_s=1$~$T$ (saturation
magnetization)  and exchange constant  $A=1.3 \times  10^{-11}$~$J/m$ are
used.  The  sample volume is  discretized into irregular tetrahedrons  with cell
size of about 3~$nm$. The damping factor is fixed to 0.02.

\begin{figure}
\begin{center}
\includegraphics[width=3in]{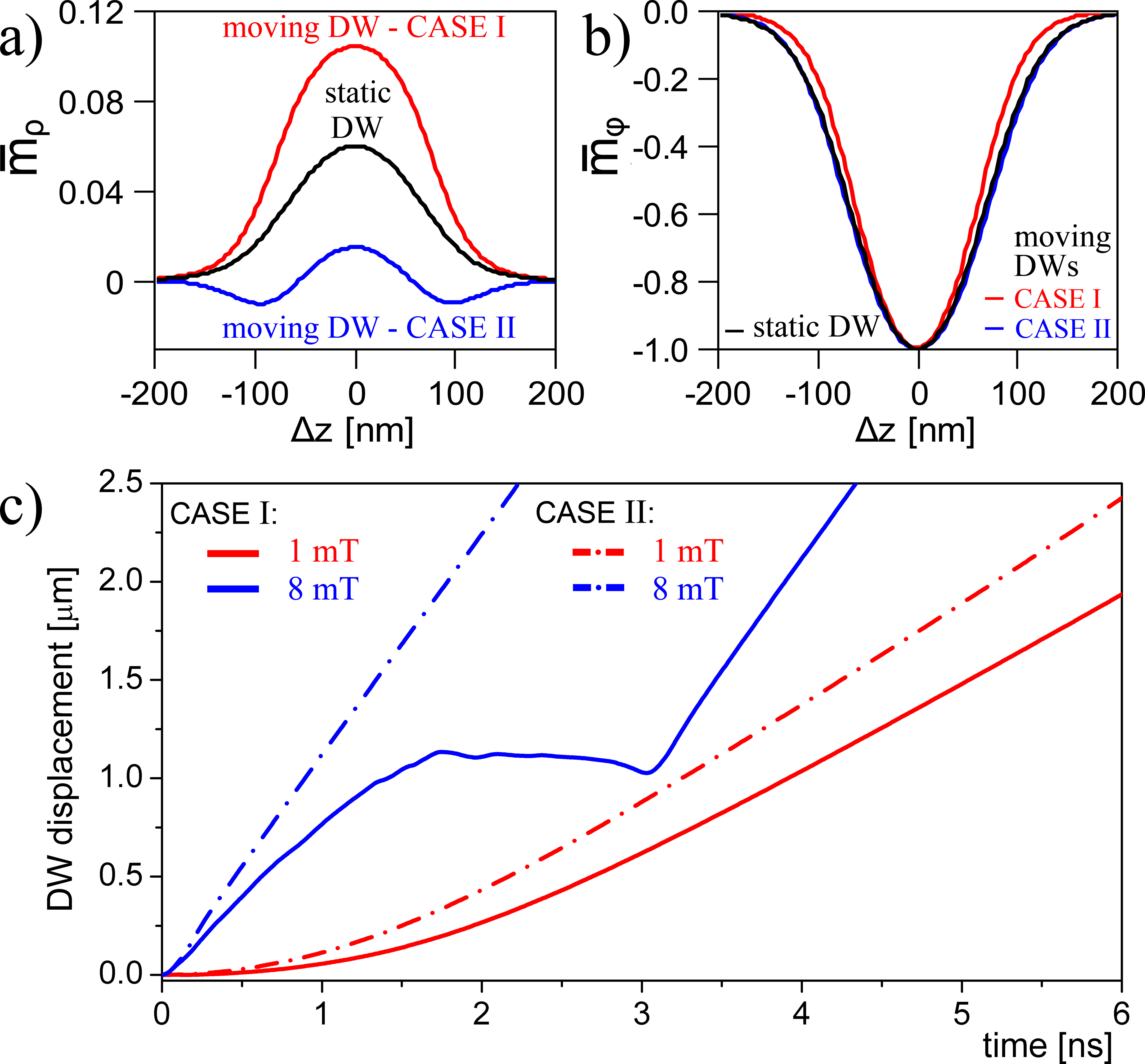}
\end{center}
\caption{\label{fig2} Comparison of the configuration between a static
DW and moving  DWs in both case I and  II by plotting $\bar{m}_{\rho}$
and $\bar{m}_{\phi}$ in  (a) and (b) respectively. The  moving DWs are
driven by  a 1 $mT$ field  in opposite directions  (+$z$ corresponding to
case I  and -$z$ to case II).  (c) DW displacement versus  time in the two
cases driven by two different fields.}
\end{figure}

To drive the DW, a  magnetic field is applied along the tube. In
flat  strips,  changing  the  orientation  of the  DW  or  the  field
direction  (left or  right) does  not affect  the DW  dynamics.  It is,
however, not the case in the tube.
This is because of the non-zero $m_\rho$ of the static DW. Considering
 the torque  exerted by  the field on  the DW,  which is given  by the
 cross product of  the field and magnetization, two  situations can be
 distinguished.  In one  case (I),  the torque  tends to  increase the
 already existing  $m_\rho$, while suppressing it in the other (case  II). In
 those two cases, the DW should therefore be distorted differently and
 thereby  move differently  as  well.  This  symmetry break is  indeed
 observed in the simulations  of the DW motion driven  by fields applied in
 opposite  directions.   Given  the  vorticity  of  the  DW  shown  in
 Fig.~\ref{fig1}, the field in +$z$  corresponds to case I and -$z$ to
 case II.  Figure.~\ref{fig2}a  shows  the expected  enhancement  and
 compression of $\bar{m}_{\rho}$  for the moving DW in  case I and II
 respectively. The symmetry-break effect is already remarkable for a 1
 $mT$ applied  field. The  plot of $\bar{m}_{\phi}$  in Fig.~\ref{fig2}b
 shows  that the  moving DW  width  is also  different in  case I  and
 II. The  displacement of the DW as  a function of time  is plotted in
 Fig.~\ref{fig2}c for two  field values (1 and 8 $mT$).  In case II, the
 DW always reaches a constant speed after a period of acceleration. In
 case I, the motion is more complex.  At low fields, the DW reaches a
 constant speed,  which is always lower  than that in case  II for the
 same field value. This velocity difference can be attributed to the different DW width during its motion in case I and II as shown in Fig.~\ref{fig2}b,  
 because the DW velocity is proportional  to the DW  width \cite{walker}.   Above a
 critical  field  (about  5  $mT$),  the motion  shows  a  'three-stage'
 behavior.   In  the  first  stage,  the DW  moves  with  a  constant
 speed. In  the second  stage, the average  velocity of the  DW becomes
 close to  zero. In  the third  stage, the DW  resumes its  motion and
 acquires a higher speed compared to the first stage.  Notice that the
 DW speed of the third stage is exactly the same as that in
 case II for the same field value.

\begin{figure}
\begin{center}
\includegraphics[width=2.8in]{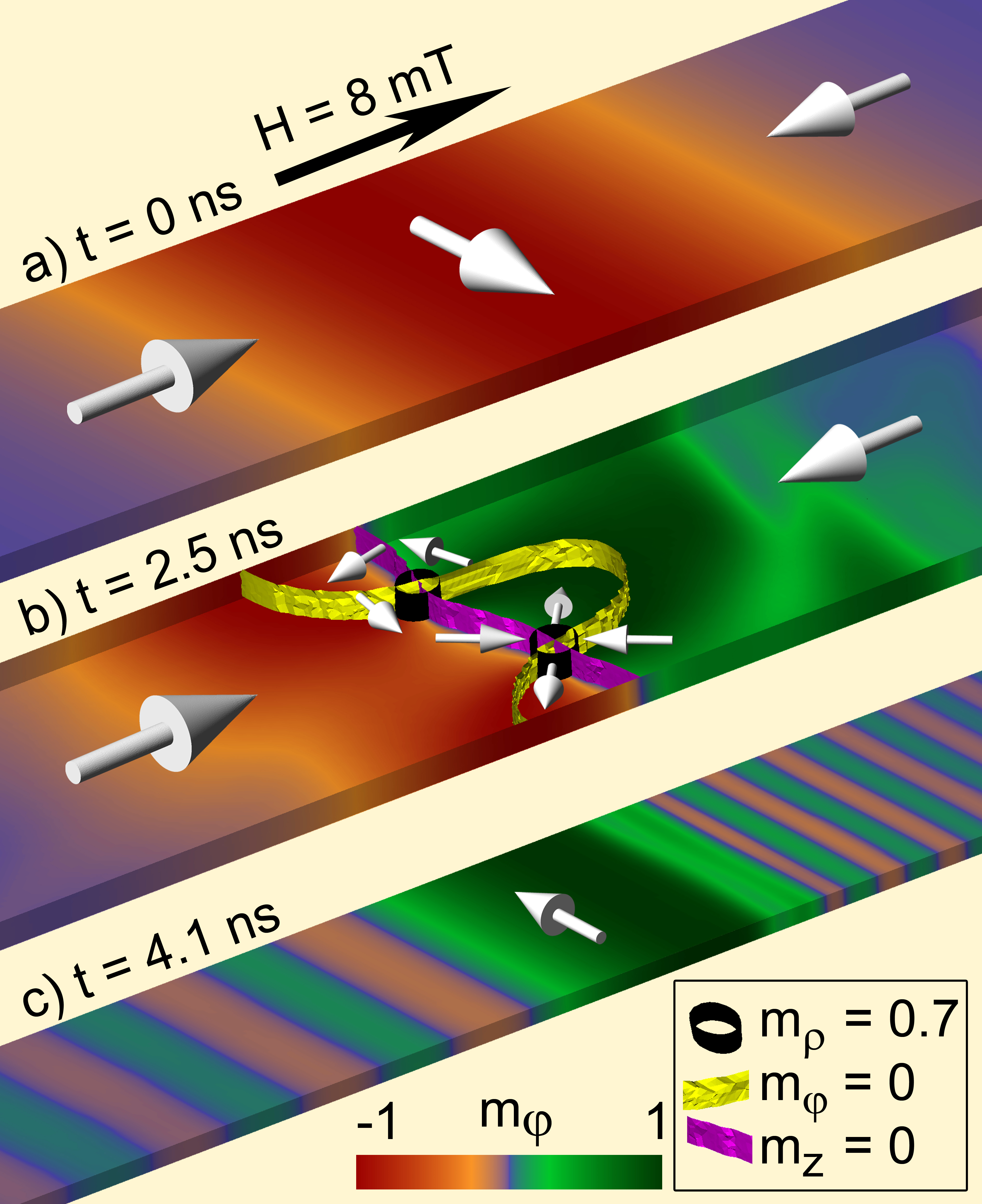}
\end{center}
\caption{\label{fig3} Snapshots of  the DW configuration before, during and after a breakdown occurring  in case I. For  better visualization, the  flat strip after
'unrolling' is shown. The magnetization is indicated by both the color
and  white arrows. In  (b), only  the bottom  surface and  the lateral
edges of the  strip are shown. Three isosurfaces  are utilized to locate
the vortex-antivortex cores. In (c), the view of the DW is zoomed out to show the spin-wave tails.}

\end{figure}

 The 'three-stage' motion of the DW in case I is caused by a breakdown
 process,  which  is   demonstrated  in  Fig.\ref{fig3}.   For  better
 visualization,  the 'unrolled'  tube is shown as  a flat  strip.  As
 mentioned before, in  case I, the initial $\bar{m}_{\rho}$  of the DW
 is further increased by the  field torque.  At a critical field, this
 distortion (large radial  component of the magnetization) leads
 to the collapse of the DW. This breakdown process is characterized by
 the  nucleation   of  a  vortex-antivortex  pair.    The  vortex  and
 anti-vortex cores  are indicated by the crossings  of two isosurfaces
 ($m_z=0$   and    $m_\phi=0$)   and   isosurface    $m_\rho=0.7$   in
 Fig.~\ref{fig3}b.   After the pair  is created,  they move  away from
 each other and eventually meet on the other side  of the tube and
 annihilate. During this  process, the DW motion is  very complex. The
 DW stops to  move or even moves backward  momentarily.  This velocity
 drop is similar to the Walker breakdown occurring in  flat strips.   Notice that
 after the annihilation of the  pair, the DW reverses its vorticity as
 shown  in Fig.~\ref{fig3}c  and its  motion thereafter  switches from
 case I to  II.  This breakdown process in the  tube differs from that
 in flat strips  mainly in two aspects.  First,  the breakdown in the
 tube involves a vortex-antivortex pair  instead of just a single vortex or
 anti-vortex as in flat strips \cite{nakatani}.  This is attributed to
 the  lack of  lateral boundaries  in the  tube and  the  requirement to
 conserve the  winding number. A similar  process of vortex-antivortex
 pair  creation and annihilation  has been  well understood  in vortex
 dynamics  \cite{hertel_prl06,   hertel_prl07}.   Because  the  energy
 needed to  create a vortex-antivortex pair  is higher than  that for a
 single vortex/anti-vortex, the breakdown in tubes should have a higher
 threshold than  that in flat strips. Our  simulations confirmed this
 (data not shown  in this paper).  Secondly, the  breakdown process in
 the tube  is not repetitious as in  thin strips.  This is  due to the
 symmetry-break of the DW motion in the tube as discussed before.

\begin{figure}
\begin{center}
\includegraphics[width=3in]{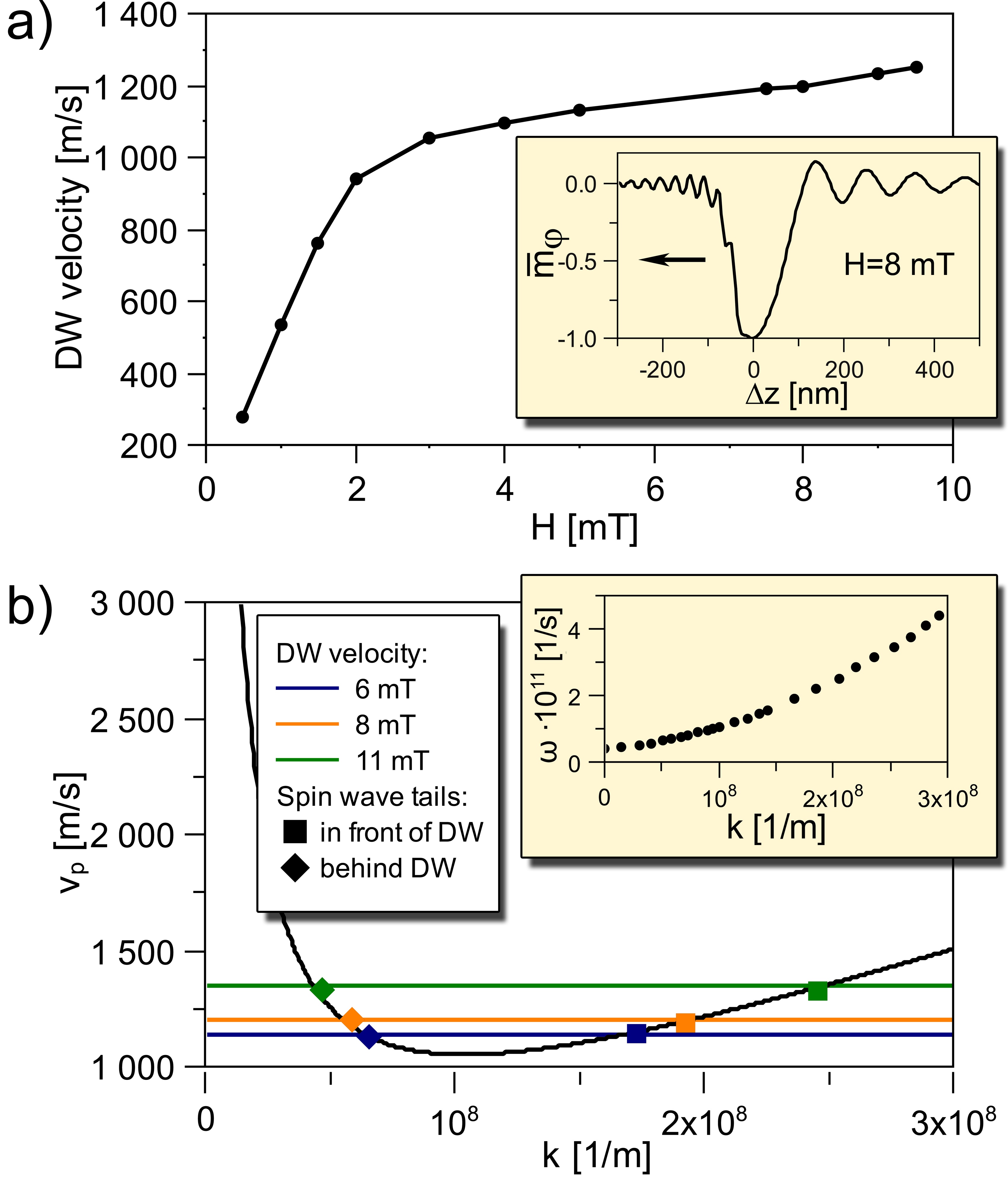}
\end{center}
\caption{\label{fig4}  (a)  DW  mobility  of  the  nanotube  in  case
II.  Inset: A  snapshot of  the  moving DW  configuration by  plotting
$\bar{m}_{\phi}$ near  the DW. The arrow  indicates the moving direction
of the DW  driven by a 8 $mT$  field. (b) Spin-wave phase  velocity as a
function of wave vector extracted  from its dispersion (inset). Spin-wave  tails of  the  moving DWs  driven by  three
different fields are compared to corresponding eigen modes.}

\end{figure}

We now focus  on the DW motion  in case II. In this  case, the initial
$\bar{m}_{\rho}$   of   the   DW    is   suppressed   by   the   field
torque.  Therefore,  the occurring  of  the  breakdown  triggered by  a
vortex-antivortex  pair  creation is  expected  to  be more  difficult
compared to case  I. In fact, the breakdown is  never observed in case
II. At  each field  value in our  studied range,  the DW moves  with a
nearly constant speed.  The DW velocity as a function of field
is  plotted in  Fig.~\ref{fig4}a. At  low  fields, the  DW velocity  is
linearly  dependent on  the field.  Above a  critical  velocity (about
1000 $m/s$), the DW  velocity curve shows a dramatic  change of its slope.
This mobility change  is found to be resulting  from a direct emission
of spin waves by the DW.  A typical snapshot of the spin waves
is shown in Fig.~\ref{fig3}c and the inset of Fig.~\ref{fig4}a.  One can see that spin waves are emitted
both  in  front  of and  behind the DW.  Both of  them  have
well-defined  yet different  wave  length. This characteristic manner of spin-wave emission is different from that of the Walker breakdown, in which spin waves with a broad spectrum are excited by the abrupt change of the DW structure.
 The  DW structure  in this case remains nearly  the  same except for the spin-wave tails.
   The  whole
structure  is  non-dispersive  and  moves  in  a  dynamic  equilibrium
\cite{movie}.   From topological  point  of  view, this  moving  DW is  a
supermagnonic  soliton  with  an  asymmetric  spatial  structure.   For a further
understanding, we  numerically calculated  the spin-wave   dispersion   of  the   tube   as   shown   in  the   inset   of
Fig.~\ref{fig4}b. In principle, the presence of a DW and the
external field influence the dispersion. Because of the relatively
small field applied in our study,  the field  effect on  the dispersion  is neglected. A DW can cause a phase shift to spin waves passing through \cite{hertel_prl_2004, gonzalez}, which is not important for our purpose.    The phase
velocity ($v_p$) of  the spin waves shown in  Fig.~\ref{fig4}b is then
extracted from the dispersion.  One  immediately sees that $v_p$ has a
minimum  around  1000 $m/s$, which coincides with  the critical  velocity
of the DW to emit spin waves. In
case I, the breakdown occurs  at a critical velocity around 800 $m/s$, which is less than this minimum spin-wave velocity.
One  also sees  that there  exist two  spin-wave  modes with  the same
$v_p$.  For  a supermagnonic DW with a  certain velocity $v$, its  two spin-wave
tails are found  to originate from the two  eigen modes sharing
the same  $v_p$ that is equal to $v$.  This is  shown in Fig.~\ref{fig4}b by  the excellent
match between  the spin-wave tails of  the DW and the  corresponding spin-wave eigen
modes at three different field  values.  The wave length and frequency
of the spin-wave  tails are measured from the  simulations. Because $v_p$ of the spin-wave tails is the same as the DW velocity, the spin waves have zero frequency in the moving frame of the DW\cite{movie}. The spin waves are therefore emitted because of soft-mode instability\cite{bouzidi}, which are not the 'wake' of a moving DW proposed in another DW motion damping mechanism \cite{wieser}.   Note that after the whole structure reaches a dynamic equilibrium,
the  spin-wave  tails  with  relatively  short wave  length  and  high
frequency always appear in front  of the DW.  This is due to the dispersion of the wave package occurring in its initial  stage because of the different group velocity of the two spin-wave modes. The emission of spin  waves transfers
energy stored in the DW and therefore hinders the further distortion of the
DW. Consequently,  the DW  mobility is  reduced.
 We point  out  that the
dynamics of the  DW in nanotubes is not thickness
specific.   Tubes with three  different inner  diameters, 40,  30, and
10 $nm$, have been studied and the same physics is revealed.


To conclude, we numerically demonstrate that a magnonic barrier of DW propagation in ferromagnetic nanostructures can occur in a tubular geometry. This is indicated by the significant reduction of the DW mobility and strong emission of spin waves by the moving DW when the DW velocity exceeds the spin-wave velocity. The characteristic manner of spin-wave emission is understood in terms of its dispersion. In addition, the curvature of the tube also causes the break of the left-right symmetry of the DW propagation. Since  fabrication  of   such  ferromagnetic
nanotubes is feasible \cite{nielsch_jap, nielsch_nano}, it is promising to experimentally verify these findings.

\end{document}